\documentclass[11pt]{article}
\usepackage{UF_FRED_paper_style}
\usepackage{xcolor}
\usepackage{lipsum}  
\usepackage{amsmath}
\usepackage{comment}

\newcommand{\be}{\begin{equation}}
\newcommand{\ee}{\end{equation}}
\newcommand{\bea}{\begin{eqnarray}}
\newcommand{\eea}{\end{eqnarray}}

\title{A multi-asset, agent-based approach applied to DeFi lending protocol modelling
}

\author{Amit Chaudhary 
     \thanks{a.chaudhary.1@warwick.ac.uk / amit@polygon.technology} 
\and Daniele Pinna
     \thanks{phys2172@ox.ac.uk / daniele@0vix.com} 
    }

\begin{document}
{\setstretch{.8}
\maketitle
\begin{abstract}

We assess the market risk of the DeFi lending protocols using a multi-asset agent-based model to simulate ensembles of users subject to price-driven liquidation risk. Our multi-asset methodology shows that the protocol's systemic risk is small under stress and that enough collateral is always present to underwrite active loans. Our simulations use a wide variety of historical data to model market volatility and run the agent-based simulation to show that even if all the assets like ETH, BTC and MATIC increase their hourly volatility by more than ten times, the protocol carries less than 0.1\% default risk given suggested protocol parameter values for liquidation loan-to-value ratio and liquidation incentives. 


\noindent
\noindent

\end{abstract}
}


\section{Introduction}

DeFi lending protocols have seen a significant flow of capital. The lending system's stability will depend on the collateral value that the borrowers provide. At any point in time, the system must have adequate capital to become solvent. Recently \footnote{\cite{kao2020analysis}}, research has attempted to estimate the financial risk of lending protocols associated with asset price fluctuations using Agent-based simulations. However, examples assume only two assets (one being the numeraire) are supplied and borrowed in the individual lending market. In reality, users can supply multiple assets to the lending market and borrow multiple assets. Sometimes the same asset is both borrowed and lent to capitalise on temporary incentive mechanisms aimed at attracting liquidity into the lending market. This paper presents an enhanced multi-asset model where real-time liquidation calls are executed as a result of price turbulence and borrowers face periods where they need to raise cash to remain within the tolerance limit of the protocol parameters like collateral ratio. As a case study,  we model these dynamics on the 0VIX\footnote{0VIX is the decentralized, Polygon blockchain-based open-source lending and borrowing protocol enhanced with veTokenomics, interest rate optimization curve beta , and DAO Treasury management.} lending protocol.

We show how one can ensure the lending market's resilience to adverse shocks even when multiple assets become highly volatile simultaneously. This is done by exploring portions of the phase space of 0VIX's asset-specific parameters and optimising them by requiring that over-collateralisation is retained across a wide range of simulated price volatilities while minimising the liquidation penalties to individual users. Analyses such as that the one presented here can be performed periodically on a running basis to offer individual users key insights into the risk of their portfolio positions, as well as propose re-calibrations of protocol parameters for discerning governance participants. We believe this offers investors the confidence to participate with significant capital in 0VIX lending pools.  

Our model is motivated by the fact that a multi-asset portfolio can withstand the risks associated with lending protocols. The shock in the price of a single asset can amplify the risk in another pool. The trajectory taken by multi-asset liquidations under amplified market conditions can vary over time and depends on several factors. These must all be factored into the incentives that motivate independent and profit-driven liquidators.

We assess the effect of the following factors 1) Available buy- and sell-side market liquidity across all asset pairs 2) Asset-specific liquidation incentives and maximal liquidation size offered to liquidators across all assets, and 3) Maximally allowed loan-to-value ratios for borrowing against specific assets deposited as collateral beyond which liquidations are permitted. These factors in our model contribute to the liquidation size, the collateral-loan assets chosen to be liquidated, and whether a liquidation call will be made at all.

Our model considers agents (or users) with multi-asset portfolio allocations. The users are always constrained by the individual collateral ratio of the protocol-defined Loan-to-value (LTV) ratios. The OVIX protocol stability is governed by the fact that all liabilities are redeemable. To maintain this balance, we model liquidation as an incentive mechanism where a liquidator is given an incentive to perform the foreclosure \citep{chatterjee2015quantitative} like an event. Whether it is profitable for the liquidator to perform the action is a critical feature in the stability of the system. The liquidators' profit is dependent on the traction cost of trading the bad debt in the market and the slippage cost associated with it. When the liquidator acts on the arbitrage opportunity in the bad state, the protocol benefits, decreasing the risk exposure. We stress the liquidation incentive to test the risk exposure of the protocol. The increase in liquidation incentive can reduce the systematic risk but can also disincentivise the borrowers as they see this as the potential penalty to their net borrowing cost. Therefore, protocols must optimally decide the incentive, keeping in mind the growth potential of the lending pools \citep{leshner2019compound}.

We assume that the strategic interaction between the liquidators are not present, and they are risk-averse \citep{schied2009risk}. This assumption represents the observed behaviour of liquidators where they immediately sell the debt in the market, assuming the risk of waiting is high enough to gain from any future price movements. 

We model the slippage cost to provide real-life market conditions across crypto assets \citep{makarov2020trading}. Market impact models have been extensively studied in which liquidity and volatility are critical drivers of the execution cost \citep{toth2011anomalous}. Since we model the multi-asset model and include relatively lower liquidity assets like MATIC, the slippage costs becomes vital for the market participants. We assume the functional form of the slippage model and use it to calibrate our simulations. In the future, we aim to use the recent order book depth data across Centralized exchanges and liquidity across—decentralized exchanges \citep{lehar2021decentralized} to model slippage more realistically.

Our market risk assessment of OVIX relies on an agent-based simulation that stresses the lending protocol based on highly volatile price trajectories. Financial institutions, including banks and federal reserves, have been using such techniques to ascertain the economy's financial stability~\cite{ramadiah2021agent}. The OVIX protocol will be on Polygon with significantly lower transaction costs, reducing the risk of liquidations and transactions failing due to costs. Past research has shown this bottleneck in the Ethereum Simulated EVM environment \footnote{Polygon is the leading platform for Ethereum scaling and infrastructure development. Its growing suite of products offers developers easy access to all major scaling and infrastructure solutions: L2 solutions (ZK Rollups and Optimistic Rollups), sidechains, hybrid solutions, stand-alone and enterprise chains, data availability solutions, and more. Polygon's scaling solutions have seen widespread adoption with 3000+ applications hosted, 1B+ total transactions processed, ~100M+ unique user addresses, and \$5B+ in assets secured.}

\begin{figure}[H]
    \centering
    \includegraphics[scale=0.5,width=\textwidth]{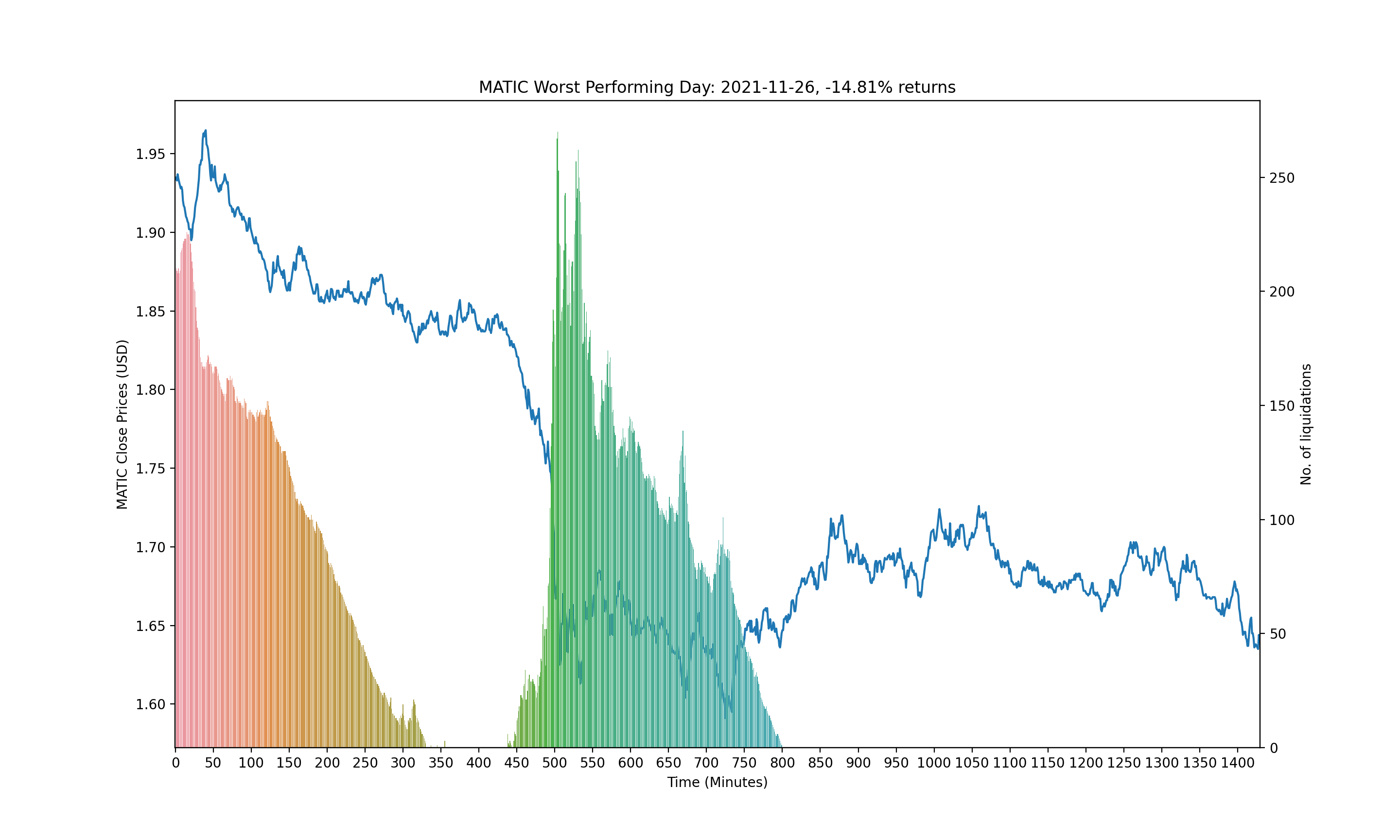}
    \caption{Simulated dual-asset MATICUSDC portfolio behavior across $1000$ long-only protocol users during the worst price drawdown day of MATICUSD history. A bar plot of liquidation events overlays the price trajectory.}
    \label{fig:1}
\end{figure}

\begin{figure}[H]
    \centering
        \includegraphics[scale=0.6,width=0.9\textwidth]{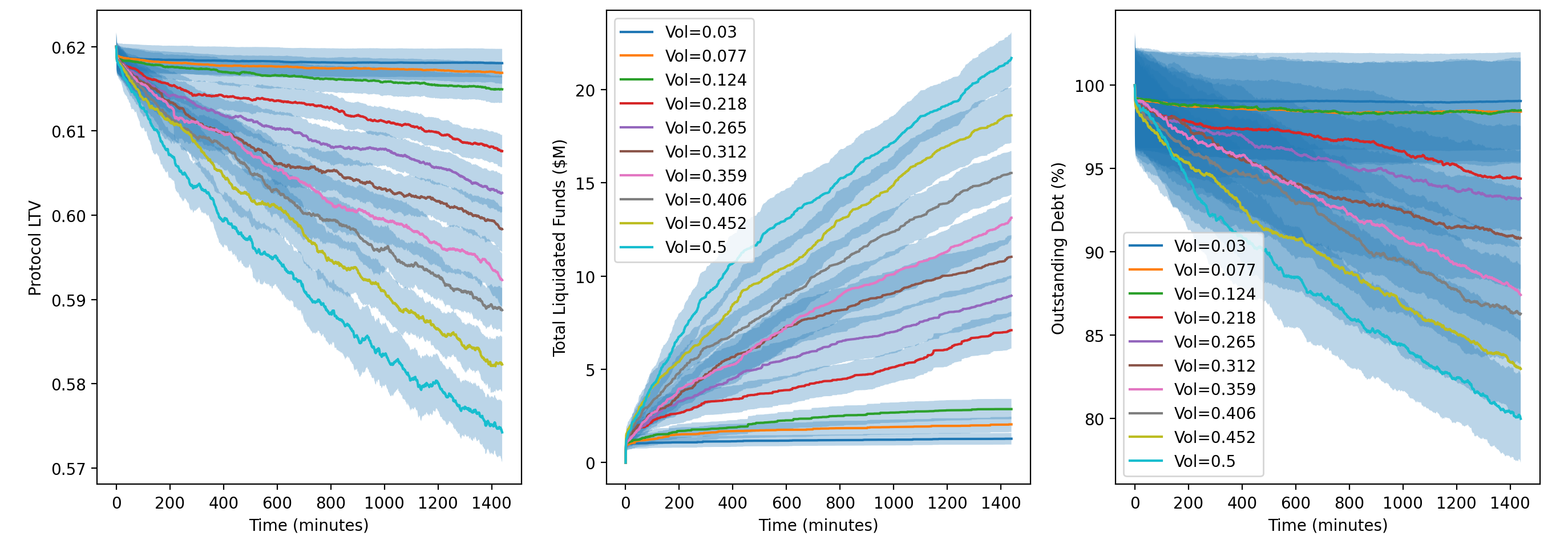}
    \caption{Simulated MATICUSDC dual-asset portfolio behavior across $1000$ protocol users subject to $1000$ randomly generated daily price trajectories with varying degree of hourly volatility (see Section~\ref{section:price_traj} for details). The shaded regions correspond to $95\%$ confidence bands. {\it Left figure:} Users' collective portfolio loan-to-value ratios over time. {\it Middle figure:} Total liquidated funds over time. {\it Right figure:} Users' collective outstanding debt as a percentage of their initial allocation.}
    \label{fig:3}
\end{figure}

Our results show that the liquidation mechanism works, and the system remains stable even in the worst price history of MATIC, when it dropped 14\% in a single day (see Figs.~\ref{fig:1} and \ref{fig:3}). Stressing the volatility of the assets has shown that the system remains within the safe LTV zone and can be scaled from a simulated \$100 million to ten times without any significant rise in the solvency of OVIX (the percentage of undercollateralized users stays well below $\%$). We also show that current protocols parameters are sufficient to face any unprecedented fall in asset prices. We test the OVIX stability across wide ranges of market volatility conditions and multiple collateral factors. We also test the robustness of liquidation incentives in the protocol and how the current incentives are sufficient to maintain the liquidity profiles in the pools where borrowers are optionally liquidated if they cross the protocol's LTV thresholds. In particular, we verify a theoretical scaling between liquidation incentives and LTV ratios above which users become likely trapped into runaway LTV factors when subject to liquidations (see Fig.~\ref{fig:constraint}). This must be avoided at all costs as it may ultimately lead to undercollateralized protocols.

\begin{figure}[H]
    \centering
        \includegraphics[width=0.87\textwidth]{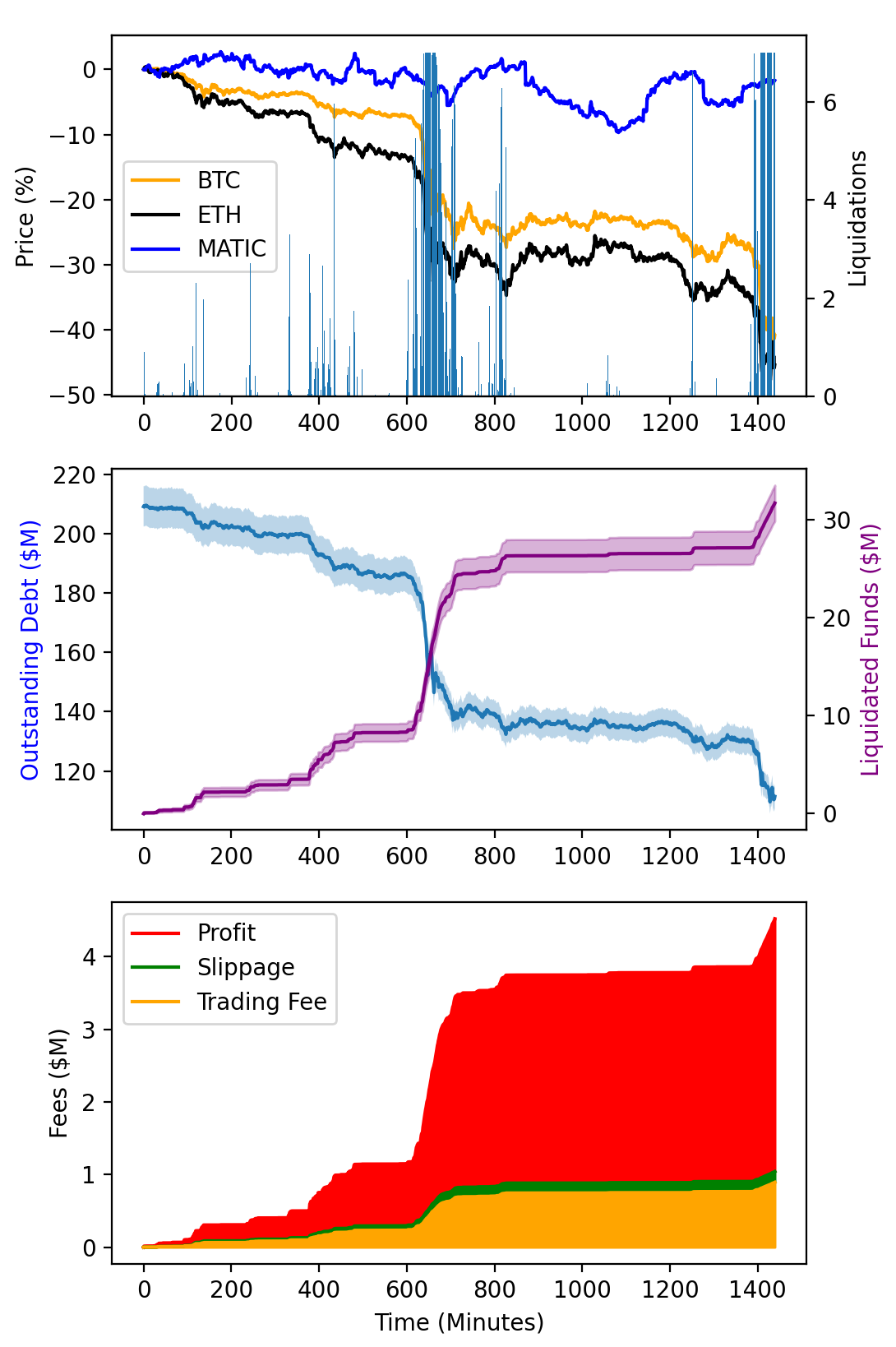}
    \caption{Simulated portfolio behavior across $1000$ protocol users throughout the `COVID Crash' day (20 Feb. 2020). User portfolios are initialized by randomly assigning collateral/loan position values (ETH, BTC, MATIC, and USDC assets were considered) consistent with a target loan-to-value ratio and portfolio size drawn from log-normal distributions for each of them (see Section~\ref{section:user_init}). $100$ such simulations are then repeated (with a different initialization each time) to collect average statistics. All figures are plotted versus time in minutes. {\it Top figure:} Liquidation events vs price drawdowns. {\it Middle figure:} Dual axis plot of users' collective outstanding debt and total liquidated funds (shaded areas represent $95\%$ confidence intervals). {\it Bottom figure:} Total liquidator profits, trading, and slippage fees.}
    \label{fig:4}
\end{figure}

\begin{figure}[H]
    \centering
        \includegraphics[width=0.87\textwidth]{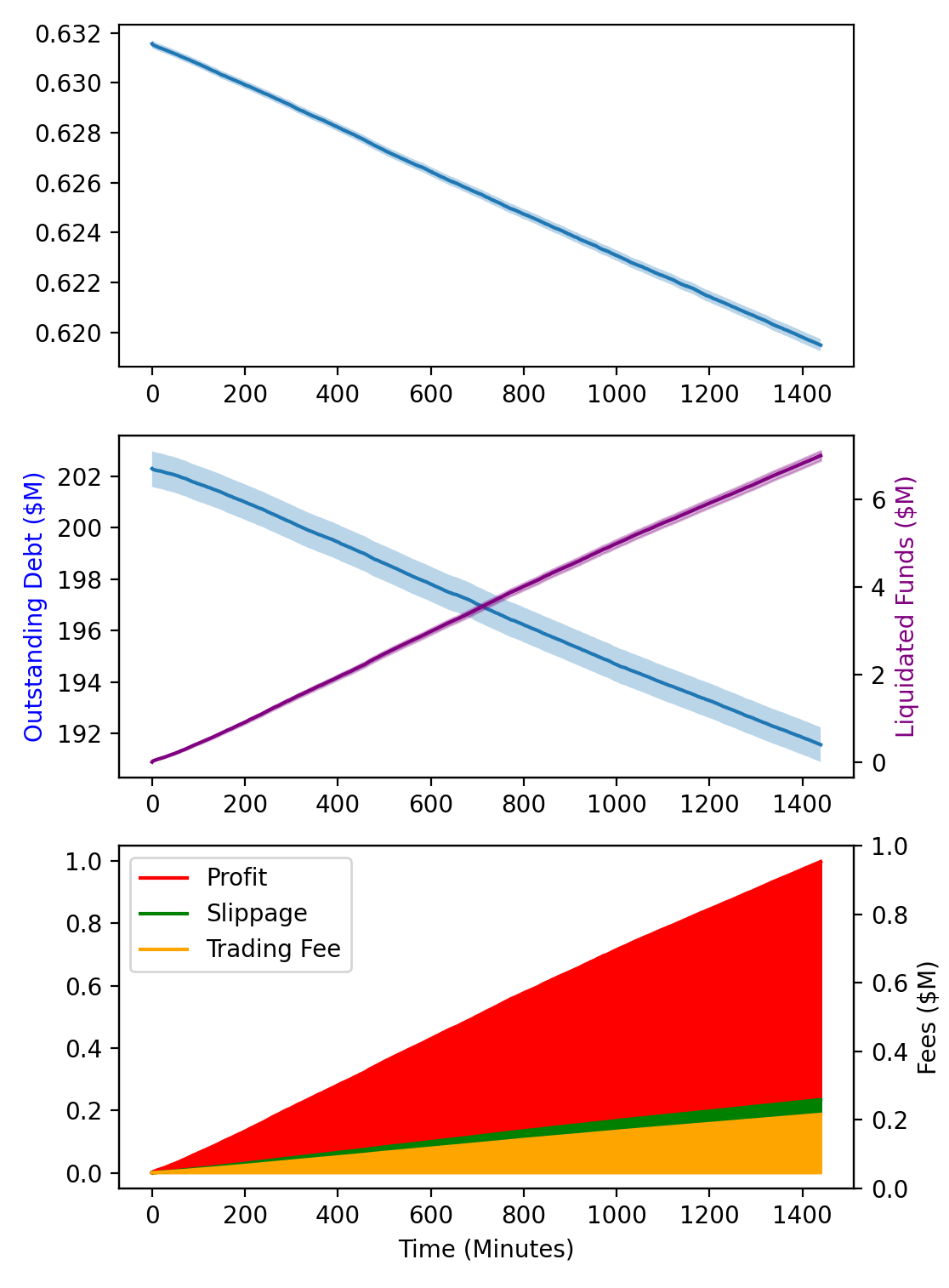}
    \caption{Simulated multi-asset portfolio behavior across $1000$ protocol users. Statistics were gathered by simulating $100$ distinct initial portfolio allocations across $100$ different price trajectories ($10,000$ distinct price trajectories in total). User portfolios are initialized by randomly assigning collateral/loan position values (ETH, BTC, MATIC, and USDC assets were considered) consistent with a target loan-to-value ratio and portfolio size drawn from log-normal distributions for each of them (see Section~\ref{section:user_init}). $100$ such simulations are then repeated (with a different initialization each time) to collect average statistics. All figures are plotted versus time in minutes. Results shown represent the average expected behavior of protocol health across any random day given the initial portfolio sized and preferred LTVs of its users. {\it Top figure:} Liquidation events vs price drawdowns. {\it Middle figure:} Dual axis plot of users' collective outstanding debt and total liquidated funds (shaded areas represent $95\%$ confidence intervals). {\it Bottom figure:} Total liquidator profits, trading, and slippage fees.}
    \label{fig:backtest}
\end{figure}

We also use Covid-pandemic time asset prices data to estimate and stress the OVIX protocol to find the likelihood of failure. The widespread economic damage caused by the COVID-19 pandemic provides a major test of the real recent stresses on crypto assets and the financial system. Our study uses the COVID data to run a market risk assessment on capital and liquidity requirements (see Fig.~\ref{fig:4}). These are then compared with similar results obtained by simulating $10,000$ distinct price trajectories across $100$ different protocol portfolios to obtain the average expectation for stress on the protocol (see Fig.~\ref{fig:backtest}).

The remainder of this study is organized as follows. Section 2 elaborates the model of the market risk framework and protocol dynamics. Section 3 presents the data and agent-based simulation framework. Section 4 contains the results, and Section 5  concludes.

\section{Model}

\subsection{Assets and Users}

We consider a set of users $N_U$ participating on the platform, and $N_A$ assets which users can deposit as collateral or borrow against other assets they have already deposited as collateral. At any given moment in time, each user $k$ is fully determined by the collateral $c^k_i$ and loan $l^k_i$ amounts (in numeraire units) of each asset $i$ they own. The total portfolio size and loan-to-value ratio (LTV) of their account can be derived as:

\bea 
\textrm{portfolio}_k(t) &=& \sum_{i=0}^{N_A}\left(c^k_i-l^k_i\right)\\
\textrm{LTV}_k(t) &=& \frac{\sum_{i=0}^{N_A}l^k_i}{\sum_{i=0}^{N_A}c^k_i}, \nonumber
\eea
where we omit the temporal dependence $(t)$ from the $rhs$ for ease of legibility. 

Each asset $i$ on the protocol is defined by liquidation LTV $\textrm{liq}^{\textrm{LTV}}_i$ beyond which any loans using the asset alone as collateral become liable for liquidation, and a closing factor $\textrm{close}_i<1$ denoting the maximal fraction of a loan portfolio consisting of that asset alone which can be liquidated. Given these protocol parameters $\left\{\textrm{liq}^{\textrm{LTV}}_i, \textrm{close}_i<1\right\}$, and a user's portfolio allocation $\{c^k_i,l^k_i\}$, one may derive each user's liquidation LTV:

\be
\textrm{LTV}^{\textrm{liq}}_k(t) = \frac{ \sum_{i=0}^{N_A}\;c^k_i\,\textrm{liq}^{\textrm{LTV}}_i}{\sum_{i=0}^{N_A}\;c^k_i}.\nonumber
\ee

Whenever a user's $\textrm{LTV}_k>\textrm{LTV}^{\textrm{liq}}_k$ a user becomes liquidatable.

Given a price trajectory for the model's assets, all the user portfolios and LTV values can be updated at each price tick by modifying the value of all the assets in their portfolio. This defines all further available actions on the protocol until the next price update. 

\subsection{Liquidators}
Whenever a user's $\textrm{LTV}_k>\textrm{LTV}^{\textrm{liq}}_k$, a liquidator may attempt to liquidate an amount $p^k_i<\textrm{close}_i\cdot l^k_i=l^k_i/2$ across one of any of the user's loan assets\footnote{Where we set $\textrm{close}_i=1/2$ across all assets for simplicity given that 0VIX currently does this equally across all assets.}. For a liquidation call to be executed, liquidators first repay some amount $p^k_i$ of the user's loan for which the protocol allows them to repossess up to $a^k_j\leq (1+inc_j)\cdot p^k_i$ of some collateral asset $j$ of their choosing among the liquidated user's collateral assets.\footnote{In some cases, a user may have collateral deposited across a number of assets but loans concentrated in one asset only such that no single collateral asset can cover the maximal liquidatable amount $l^k_i/2$}.

The incentive bonus $inc_j$ offered to liquidators, serves the purpose of incentivizing the active monitoring and execution of liquidation calls when necessary. The protocol assigns a percentage liquidation incentive $inc_j$ to each asset $j$ used as collateral to incentivize the liquidation of certain assets before others. Based on this, a liquidator will consider their potential gains minus any swap fees defined by the sum of three factors: transaction fees, trading fees, and slippage fees. In our treatment we ignore transaction fees since they can be absorbed into the trading fees $\gamma$ by potentially making them time-dependent.

\subsubsection{Slippage Costs}

The slippage fees are very relevant as they depend on the amount $x^k_j < a^k_j$ of a user's repossessed collateral asset $j$ being swapped to repay a liquidator's chosen $p^k_i$ relative to the available sell-side liquidity $V_{j\to i}$ in the $(j, i)$ asset pair and the normalized distribution $\rho_{j,i}(\epsilon)$ of such liquidity (as a function of deviation $\epsilon$ from quoted market price) across all accessible markets. The percent slippage fee $\sigma_{j,i}(x^k_j)$ is then defined by the following equations:

\bea
\frac{x^k_j}{V_{j\to i}} &=& \int_{0}^{\Delta P\left(x^k_j/V_{j\to i}\right)}\mathrm{d}\epsilon\;\rho_{j,i}(\epsilon) \\
\sigma_{j,i}(x^k_j) &=& \frac{V_{j\to i}}{x^k_j}\int_{0}^{\Delta P\left(x^k_j/V_{j\to i}\right)}\mathrm{d}\epsilon\;\epsilon\rho_{j,i}(\epsilon),
\eea

where the first equation self-consistently defines the total price slippage $\Delta P\left(x^k_j/V_{j\to i}\right)$ to swap $x^k_j$, and the second defines the percent loss due to slippage for this amount.

Whereas this quantity can be compiled from real-world data and used in our multi-asset model, it can be also effectively modelled for simulation purposes using $\omega$-polynomial approximations depending on a slippage factor $s_{\omega}$ (to be fit alongside $\omega$) from historical data:

\be
\label{slippagemodel}
\sigma_{j,i}(x^k_j) \approx \gamma + s_{\omega}\cdot  \left(\frac{x^k_j}{V_{j\to i}}\right)^\omega
\ee

From this one can succinctly write the total profit a liquidator can make for swapping an amount $x^k_j$ to reimburse themselves for liquidating an amount $p^k_i$ of user's loan asset $i$ as:

\be
\label{liquidatorprofit}
\textrm{liquidatorProfit}_{j,i}(p^k_i) = (1+inc_j)\cdot p^k_i-x^k_j(p^k_i),
\ee
where $x^k_j(p^k_i)$ is obtained by inverting the slippage equation :

\be
p^k_i= \left[1-\sigma_{j,i}(x^k_j)\right]x^k_j.
\ee

\subsubsection{Liquidation Logic}

Recently, it has been argued that real-world slippage behavior is suitably modelled in (\ref{slippagemodel}) by setting $\omega=1$  \citep{kao2020analysis}. Under such an approximation, $x^k_j$ can be explicitly solved as a function of $p^k_i$ giving:

\be
\label{amount_to_swap}
x^k_j(p^k_i) = \delta_{i,j}p^k_i+(1-\delta_{i,j})\frac{1-\gamma}{2\tilde{s}_{j,i}}\left[1-\sqrt{1-\frac{4\tilde{s}_{j,i} p^k_i}{(1-\gamma)^2}}\right]
\ee

where $\delta_{i,j}$ is the Kronecker delta, and $\tilde{s}_{j,i}\equiv s_1/V_{j\to i}$ represent an array of asset parameters\footnote{Potentially time-dependent also.} in the multi-asset model. When $j\neq i$, the swappable amount is well-defined as long as the amount chosen to repay satisfies:

\be
p^k_i < \frac{(1-\gamma)^2}{4\,\tilde{s}_{j,i}}.
\ee
In principle, if this inequality is violated, the total swap fees are effectively so large that no amount $x^k_j$ can swap for $p^k_i$. In practice however, an optimal repay amount $\bar{p}^k_i$ above which the liquidator profits decrease can be established by maximizing~\ref{liquidatorprofit}:

\be
\label{swap_optimum}
\bar{p}^k_i = \frac{(1+inc_j)^2(1-\gamma)^2-1}{4\tilde{s}_{j,i}(1+i)^2} \\
\ee

Combining (\ref{amount_to_swap}), and (\ref{swap_optimum}), with the protocol imposed $p^k_i<l^k_i/2$, the liquidator's optimization strategy can be summarized as the following constrained minimization problem across all asset pairs $(i,j)$ for every liquidatable user:

\be
\left\{ \begin{aligned} 
  &\max_{i,j}& \textrm{liquidatorProfit}_{j,i}(p^k_i)\\
  &p^k_i&= \textrm{min}\left[\bar{p}^k_i, \frac{c^k_j}{1+inc_j}, l^k_i/2\right] 
\end{aligned} \right.
\ee

\subsection{Hard Parametric Constraints}

The model just described presents large amounts of complexity resulting from the many free parameters that can be included. Their interactions and effects are generally non-trivial and only analyzable through extensive simulation efforts. However, some general hard constraints can be placed on two parameters in the system: the liquidation LTV $\mathrm{LTV}^{liq}_k$, and the net fees $inc_j$ paid to liquidators. 

As discussed, when a user's LTV crosses their liquidation threshold $\mathrm{LTV}_k>\mathrm{LTV}^{liq}_k$, liquidators become incentivized to liquidate a portion of their positions to sanitize their LTV. This however requires that the LTV of the user must invariably decrease as a result of liquidations: 

\be
\mathrm{LTV}_k(t+1)=\frac{\left(\sum_{i=0}^{N_A}l^k_i\right)-p^k_i}{\left(\sum_{i=0}^{N_A}c^k_i\right)-a_k^j}<\mathrm{LTV}_k(t)
\ee

To characterize the importance of this constraint, let us extract a condition on $\mathrm{LTV}_k(t)$ and $inc_j$ simultaneously, one has:

\bea
\label{eq:LTVconstraint}
\mathrm{LTV}_k &>& \frac{\left(\sum_{i=0}^{N_A}l^k_i\right)-p^k_i}{\left(\sum_{i=0}^{N_A}c^k_i\right)-(1+inc_j)p^k_i} \nonumber\\
&=&\frac{\left(\sum_{i=0}^{N_A}l^k_i\right)/\left(\sum_{i=0}^{N_A}c^k_i\right)-(p^k_i/\left(\sum_{i=0}^{N_A}c^k_i\right))}{1-(1+inc_j)\left[p^k_i/\left(\sum_{i=0}^{N_A}c^k_i\right)\right]} \nonumber\\
&=&\frac{\mathrm{LTV}_k-\left[p^k_i/\left(\sum_{i=0}^{N_A}c^k_i\right)\right]}{1-(1+inc_j)\left[p^k_i/\left(\sum_{i=0}^{N_A}c^k_i\right)\right]} \nonumber\\
&\Longrightarrow& \mathrm{LTV}_k<\frac{1}{1+inc_j}\simeq 1-inc_j
\eea

\begin{figure}[H]
    \centering
        \includegraphics[scale=.8,width=\textwidth]{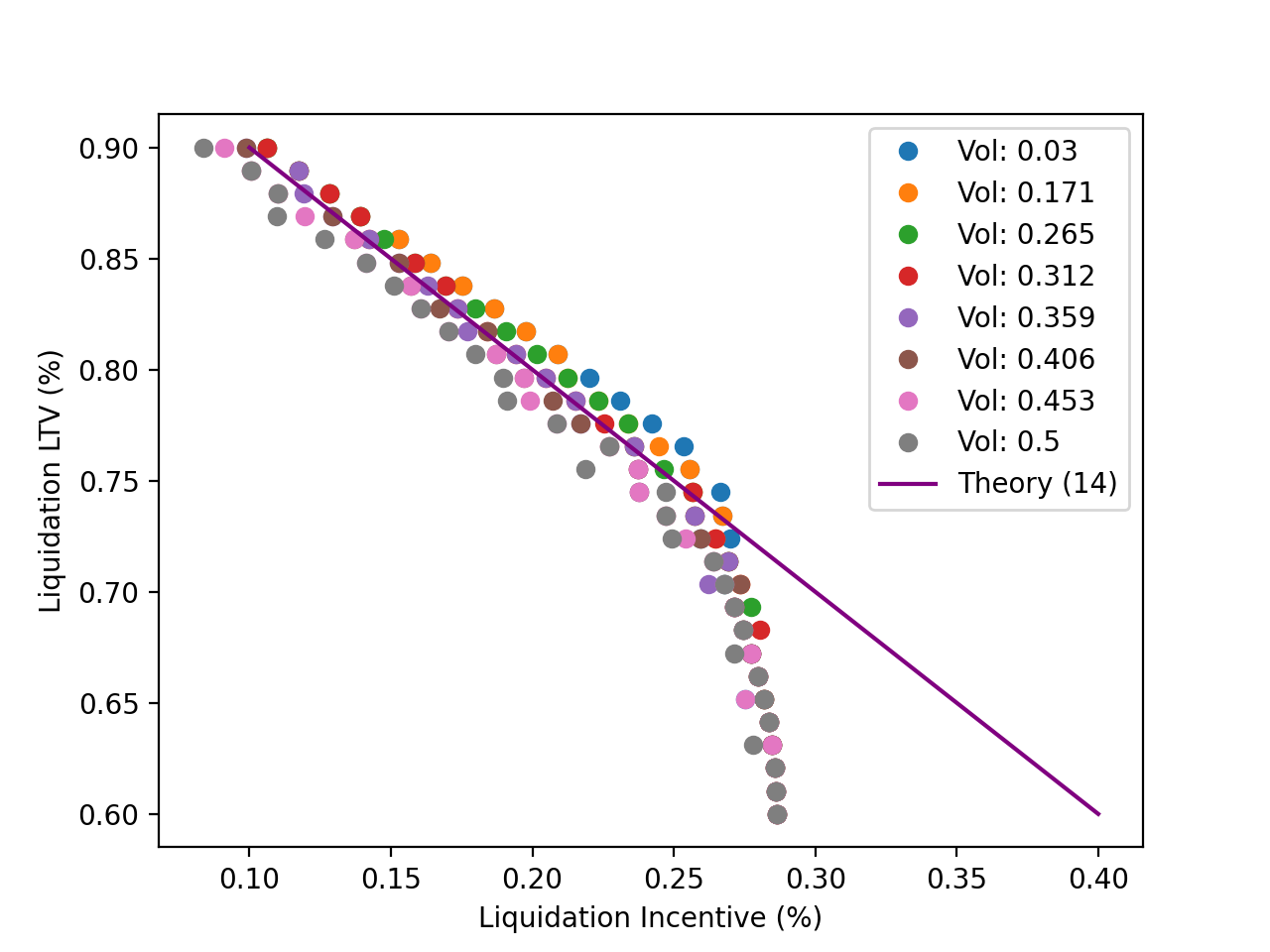}
    \caption{Undercollateralization frontier: Scatter plot of pair values (liquidation LTV, liquidation incentive) at which the number of undercollateralized users on the protocol exceeds $>1\%$. For each such pair, $1000$ MATICUSD user portfolios were simulated $100$ times each to collect sufficient statistics. The solid purple line represents the theoretical prediction based on our model discussed in the lead-up to equation (\ref{eq:LTVconstraint}). Deviations from theoretical trend at lower liquidation LTV values are due to finite lower bounds on the initial LTV of simualted users.}
    \label{fig:constraint}
\end{figure}

A relationship exists between the maximal LTV reached by a user and the liquidation incentive paid to liquidators, such that, if violated, leads to the systemic creation of undercollateralized users. Equation (\ref{eq:LTVconstraint}) is a direct prediction of our model whose verification can be seen in Fig.~\ref{fig:constraint} where deviations from trend at lower liquidation LTV values is due to the hard minimum initial LTV users were allowed to have. 

\begin{figure}[H]
    \centering
        \includegraphics[scale=.8,width=\textwidth]{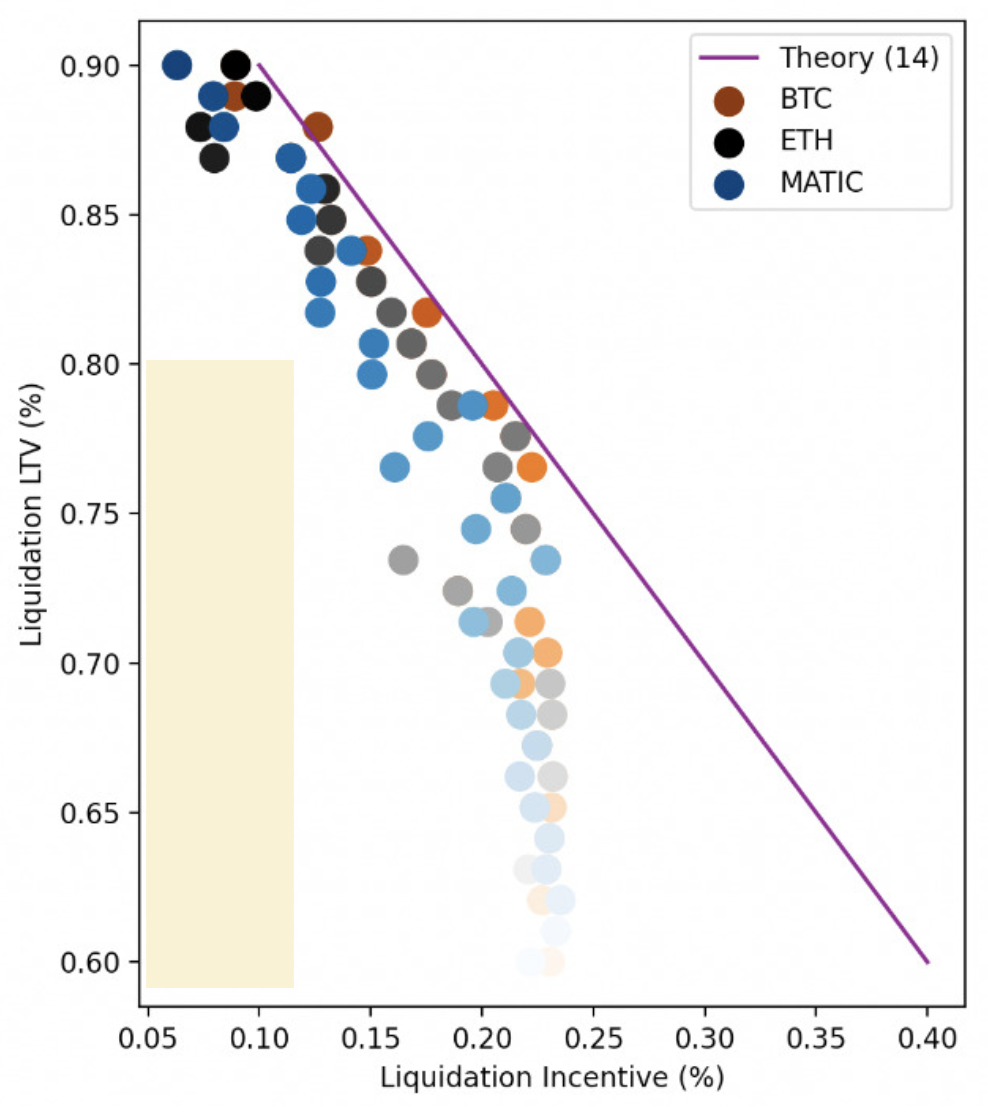}
    \caption{Scatter plot of pair values (liquidation LTV, liquidation incentive) at which the number of undercollateralized users on the protocol exceeds $>0.1\%$. Unlike Fig.~\ref{fig:constraint} here we show results for MATICUSD, ETHUSD, and BTCUSD simulated portfolios. Color shading is proportional to final average protocol LTV (darker colors = higher final LTV values). Simulations are performed on ensembles of $2000$ users with $200$ price trajectories for each (liquidation LTV, liquidation incentive) pair values to collect sufficient statistics. Solid purple line represents the theoretical undercollateralization frontier for reference (\ref{eq:LTVconstraint}). Yellow shaded area represents where optimal protocol parameters should lie.}
    \label{fig:optimization}
\end{figure}

This condition then allows one to set arbitrary bounds for the level of undercollateralization risk the protocol is willing to assume. Furthermore, protocol parameters may be optimized to minimize such risk. In Fig.~\ref{fig:optimization}, we plot the risk frontier for generating undercollateralized users with $>0.1\%$ probability. This sets hard constraints on the kind of parameter values the protocol should be advised to choose from (yellow shaded area in figure).

\section{Data and Simulations}

\subsection{User Initialization}
\label{section:user_init}

Agents in our model are considered passive. Their portfolio is randomly allocated at the beginning of the simulation and assumed to not be adjusted throughout the course o the simulation. To begin the initialization process, collateral and loan asset values $\{c^k_i, l^k_i\}_{i=0}^{N_A}$ are randomly assigned. For the results presented in this paper, we choose to neglect portfolios where users borrow and collateralize identical assets. To impose this, we unwind portfolio allocations after having generated them.  

To begin a simulation, each user is assigned a portfolio size $\textrm{portfolio}_k(0)$ and LTV value $\textrm{LTV}_k(0)$. These are drawn randomly from lognormal distributions targeting mean portfolio values of $\$ 5000$ and mean LTV values of $0.6$ (with a hard minimum LTV value of $0.45$). 

The initial portfolio asset values can then be individually rescaled $\{c^k_i, l^k_i\}_{i=0}^{N_A}\to \{r_C^k\cdot c^k_i, r_L^k\cdot l^k_i\}_{i=0}^{N_A}$ such that they respect $\left(\textrm{portfolio}_k(0), \textrm{LTV}_k(0)\right)$ assigned. The rescaling factors are given by:

\bea
r^k_C = \frac{\textrm{portfolio}_k(0)}{\left(1-\textrm{LTV}_k(0)\right)\sum_{i=0}^{N_A}c^k_i} \\
r^k_L = r^k_C\cdot\textrm{LTV}_k(0)\cdot\frac{\sum_{i=0}^{N_A}c^k_i}{\sum_{i=0}^{N_A}l^k_i}.
\eea

\begin{figure}[H]
    \centering
        \includegraphics[scale=.8,width=\textwidth]{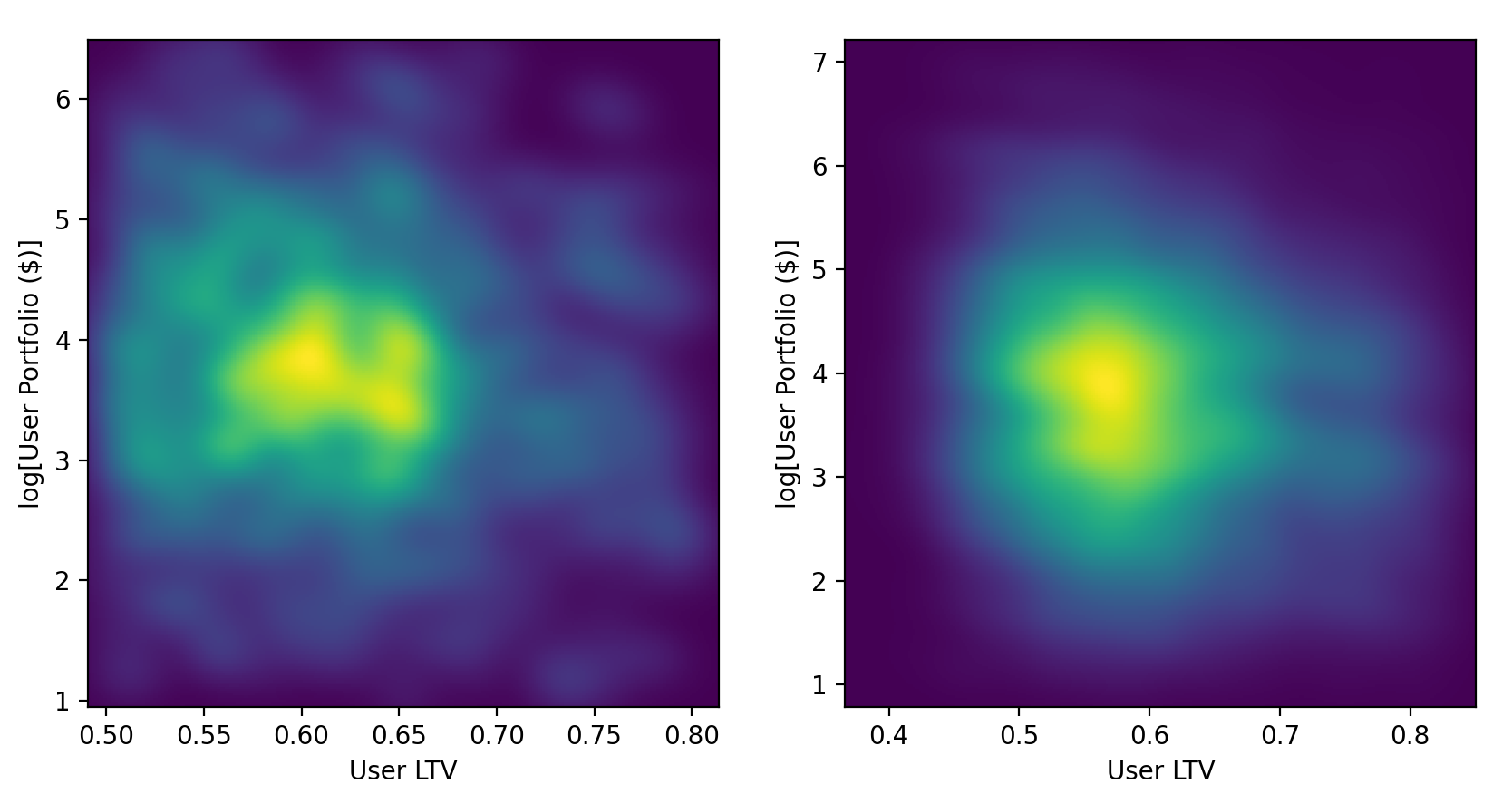}
    \caption{Initial ({\it left figure}) and final ({\it right figure}) user portfolio distributions corresponding to a single MATICUSD volatility simulation ($40\%$ volatility results are shown). Users are collectively characterized according to their loan-to-value ratios and portfolio size.}
    \label{fig:5}
\end{figure}

A typical initial portfolio ensemble is shown in the left figure of Fig.{\ref{fig:5}}

\subsection{Price Trajectories}
\label{section:price_traj}

To run our model, we have collected the past three years of price data across Bitcoin, Ethereum, MATIC, and USDC with minute tick level resolution. This allows us to both simulate our model under actual past price dynamics, and generate new cross-asset price dynamics by randomly selecting portions of the historical data. Since our agents are not allowed to modify their portfolio allocations, simulations are run with only one day's worth of price data. Simulating beyond this timescale is possible by unrealistic. To simulate specific price volatility, each randomly filtered price data sequence is individually rescaled to match the desired \emph{hourly} target volatility before commencing the simulation run.

\begin{figure}[H]
    \centering
        \includegraphics[width=0.8\textwidth]{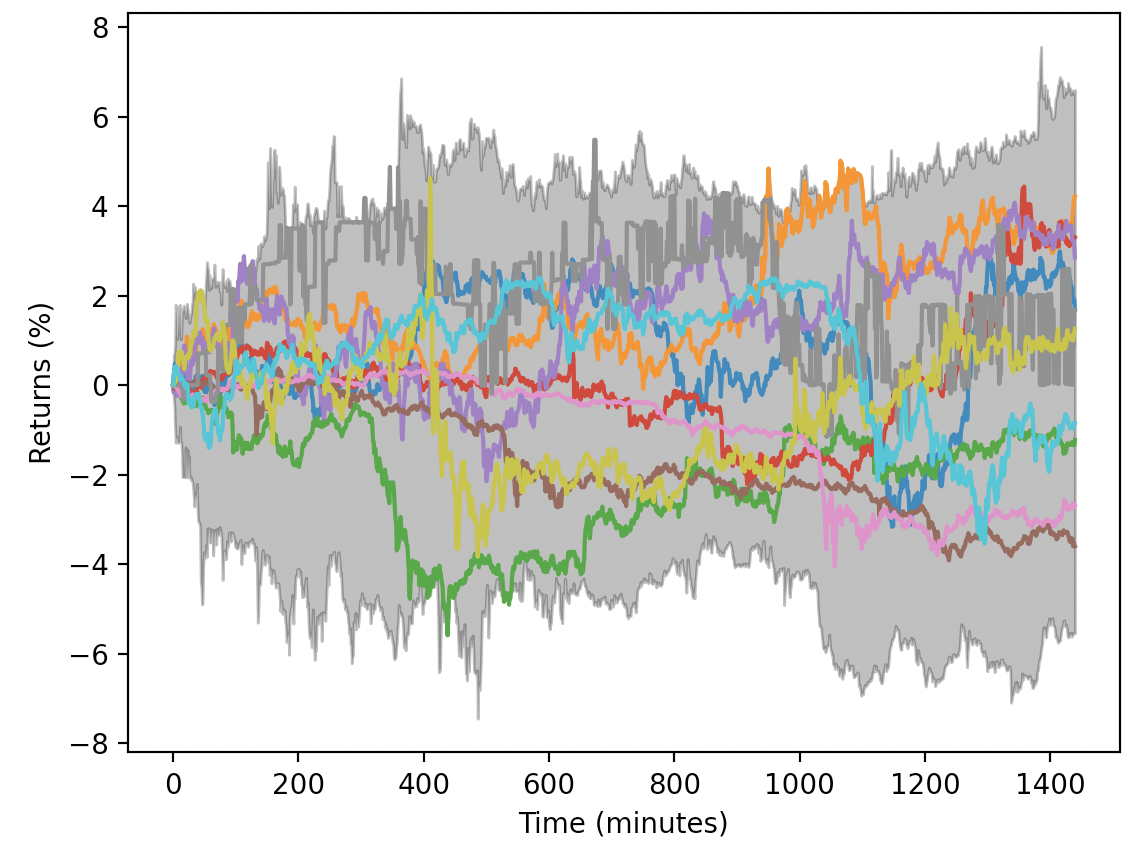}
    \caption{Example ensemble of generated price trajectories for the ETH asset. Shaded areas represent bounds of $100$ generated trajectories of which only $10$ are plotted for visual convenience.}
    \label{fig:2}
\end{figure}

\subsection{Data}

To collect statistically significant data, $1000$ simulation runs are performed for each set of desired protocol parameters. Each simulation run tracks the evolution of $1000$ user portfolios across Bitcoin, Ethereum, MATIC, and USDC regenerating their individual allocation at each run so that the results are not biased to specific portfolio ensembles. Throughout the simulations we assume that the total available sell-side market volume for swaps is $\$100\,\textrm{M}$ for all X-MATIC pairs (where 'X' stands for BTC, ETH, and USDC), and $\$1000\,\textrm{M}$ for the rest.  

\section{Results}

\section{Discussion and Conclusions}

We have built an agent-based, multi-asset simulator for lending markets which allows for extensive testing and aggregation of portfolio statistics across varying stressors and protocol parameters. We have demonstrated its principal characteristics by simulating portfolio behaviors subject to volatile real-world data such as the COVID market crash of 20 February 2020, as well as artificially generated high-volatility price scenarios. In principle, such a framework allows for a detailed protocol parameter exploration under various circumstances, allowing for their optimization. What our model does not include is the dynamical re-allocation of user portfolios during market unwinding events. We have argued for their neglect by remarking that our simulations are valid only in a limit of passive user behavior. As such, we have refrained from running our models for longer than a daily ($1440$ minute) timescale. In the future we plan to add intra-day user interactions characterized by an intrinsic user-specific timescale describing the frequency with which they may rebalance their portfolios. Similarly to the loan-to-value ratio and portfolio sizes, this too can be drawn from a suitable distribution. Overall, we believe our framework is ideally suited for decentralized lending markets such as the soon-to-be-launched 0VIX, due to the potential for generating data-driven protocol upgrade proposals that governance token holders can evaluate and vote on.

\medskip

\bibliography{references.bib} 

\newpage

\end{document}